\documentclass[conference]{IEEEtran}
\IEEEoverridecommandlockouts

\usepackage{graphicx}
\usepackage[hidelinks]{hyperref}
\usepackage{amsmath}
\usepackage{amssymb}
\usepackage{cite}
\usepackage{url}
\usepackage{booktabs}
\usepackage{amsfonts}
\usepackage{longtable}
\usepackage{nicefrac}
\usepackage{microtype}
\usepackage{xcolor}
\usepackage{enumitem}
\usepackage{comment}
\usepackage{subcaption}
\usepackage[most]{tcolorbox}
\tcbuselibrary{skins, breakable}

\usepackage{listings}
\usepackage{pdfpages}
\usepackage{multirow}
\usepackage{verbatim}
\usepackage{tabularx}
\usepackage{pifont}
\usepackage{lipsum}
\usepackage{colortbl}
\usepackage{float}
\usepackage{hyphenat}
\usepackage{array}
\usepackage{wrapfig}
\usepackage{fancyvrb}

\newcommand*\colourcheck[1]{%
  \expandafter\newcommand\csname #1check\endcsname{\textcolor{#1}{\ding{51}}}
}
\colourcheck{blue}
\colourcheck{green}
\colourcheck{red}
\colourcheck{teal}
\newcommand*\colourcross[1]{%
  \expandafter\newcommand\csname #1cross\endcsname{\textcolor{#1}{\ding{55}}}%
}
\colourcross{blue}
\colourcross{green}
\colourcross{red}
\colourcross{purple}
\colourcross{gold}

\usepackage{algorithmic}
\usepackage{textcomp}

\newtcolorbox{casebox}[1]{
    colback=white,
    colframe=blue!50!black,
    fonttitle=\bfseries,
    fontupper=\scriptsize, 
    title=#1,
    enhanced,
    drop shadow={black!50!white},
    breakable,
    boxsep=2pt,        
    enhanced,
    left=2pt, right=2pt,
    top=2pt, bottom=2pt,
    before skip=5pt,
    after skip=5pt,
    drop fuzzy shadow={black!15!white},
}

\lstdefinestyle{verilog}{
    basicstyle=\ttfamily\tiny,
    keywordstyle=\color{blue},
    commentstyle=\color{green!60!black},
    stringstyle=\color{red},
    numbers=left,
    numberstyle=\tiny\color{gray},
    stepnumber=1,
    numbersep=5pt,
    backgroundcolor=\color{white},
    frame=single,
    rulecolor=\color{black!30},
    tabsize=4,
    captionpos=b,
    breaklines=true,
    breakatwhitespace=true,
    showstringspaces=false
    aboveskip=0pt,     
    belowskip=0pt,     
    abovecaptionskip=2pt,
    belowcaptionskip=2pt,
}

\begin{document}

\title{
	\textit{VeriContaminated}:
	Assessing
	LLM-Driven {Veri}log Coding for
	{Data} {C}ontamination
}

\author{%
Zeng~Wang$^\dag$$^*$,\thanks{\textsuperscript{*}Authors contributed equally to this research.}
Minghao~Shao$^\dag$$^\ddag$$^*$,
Jitendra~Bhandari$^\dag$,
Likhitha~Mankali$^\dag$, \\
Ramesh~Karri$^\dag$,
Ozgur~Sinanoglu$^\ddag$,
Muhammad~Shafique$^\ddag$,
Johann~Knechtel$^\ddag$
\\
\IEEEauthorblockA{
$^\dag$NYU Tandon School of Engineering, USA\\
$^\ddag$NYU Abu Dhabi, UAE \\
\normalsize{Email:\{zw3464, shao.minghao, jb7410, likhitha.mankali, rkarri, ozgursin, muhammad.shafique, johann\}@nyu.edu}}
}

\maketitle

\begin{abstract}
Large Language Models (LLMs) have revolutionized code generation, achieving exceptional results on various established benchmarking frameworks.
However, concerns about data contamination---where benchmark data inadvertently leaks into pre-training or fine-tuning datasets--- raise questions about the validity of these evaluations.
While this issue is known, limiting the industrial adoption of LLM-driven software engineering, hardware coding has received little to no attention regarding these risks.
For the first time, we analyze state-of-the-art (SOTA) evaluation frameworks for Verilog code generation (VerilogEval and RTLLM), using established methods for contamination detection (CCD and Min-K\% Prob).
We cover SOTA commercial and open-source LLMs (CodeGen2.5, Minitron 4b, Mistral 7b, phi-4 mini, LLaMA-\{1,2,3.1\}, GPT-\{2,3.5,4o\}, Deepseek-Coder, and CodeQwen 1.5),
in baseline and fine-tuned models (RTLCoder and Verigen).
Our study confirms that data contamination is a critical concern. We explore mitigations and the resulting trade-offs for code quality vs fairness (i.e., reducing contamination toward unbiased benchmarking).

\end{abstract}


\section{Introduction}

Large Language Models (LLMs) like GPT-4~\cite{openai2024gpt4} and Gemini~\cite{team2023gemini} have exhibited remarkable capabilities in comprehending text semantics and generating code across different programming languages.
However, a critical challenge for evaluating these models is data contamination, which
occurs when benchmarks inadvertently include samples from the model's pre-training corpus~\cite{balloccu2024leak}.
Such a scenario would unfairly inflate performance estimates and undermine the reliability of comparative assessments.
With impressive performance metrics reported for many LLMs, the undisclosed nature of the pre-training datasets, common even among open-source models~\cite{touvron2023llama}, raises significant concerns for
such contamination~\cite{magar2022data,xu2024benchmark,ishihara-2023-training,10.1145/3523273}.

\begin{figure}[!t]
    \centering
    \includegraphics[width=1.0\columnwidth]{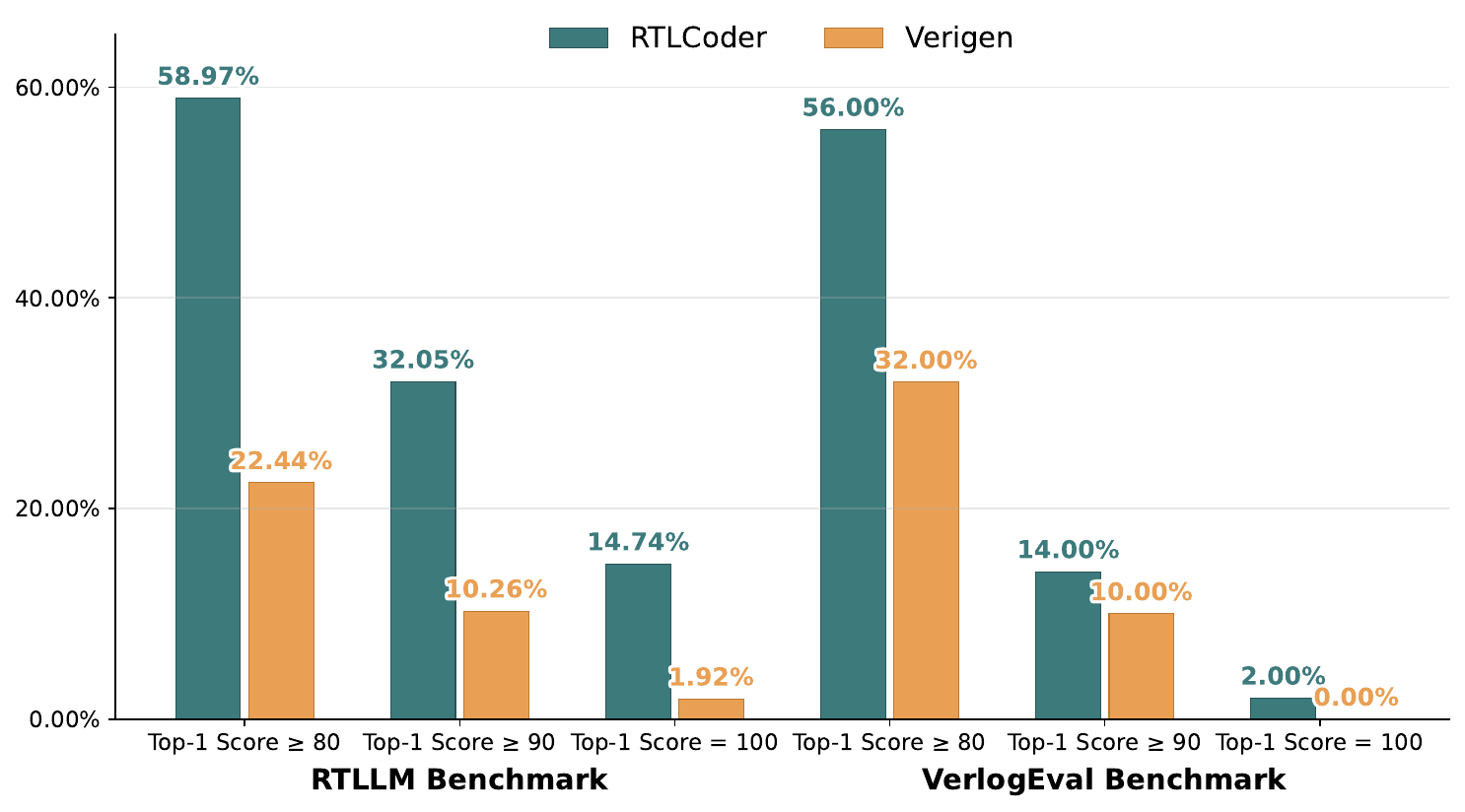} 
    \caption{Top-1 scores for Verilog Abstract Syntax Tree (AST) similarity, powered by Dolos~\cite{maertens2024discovering},
	    between
	    golden solutions and closest designs in training datasets.
	    RTLCoder and Verigen show significant similarities to the RTLLM and VerilogEval benchmarks, suggesting severe data contamination.
    }
    \label{fig:DC-motivation}
    \vspace{-2mm}
\end{figure}

LLMs have also shown considerable promise in the domain of hardware design~\cite{wang2024llms,
liu2023verilogeval,thakur2023autochip,lu2024rtllm,thakur2023verigen,
kande2023llmassisted,fang2024assertllm,
qiu2024autobench,bhandari2024llm,
wu2024chateda,liu2023chipnemo}. 
However, as in the software domain,
data contamination poses a challenge when establishing LLMs for hardware designs.
The performance and utility of these domain-specific LLMs depend on the availability of large, high-quality datasets tailored to the complexities of hardware design.
Given the relative sparsity of such data, it is easy to see that
evaluation benchmarks may (inadvertently) include some of the same samples that were part of the pre-training data.
Figure~\ref{fig:DC-motivation} confirms this hypothesis through assessment of code similarity for the state-of-the-art (SOTA) datasets of Verigen~\cite{thakur2023verigen} and RTLCoder~\cite{RTLCoder} 
against the SOTA benchmarks VerilogEval~\cite{liu2023verilogeval} and RTLLM~\cite{lu2024rtllm}.
Such large overlaps between training and testing data are highly likely to induce data contamination and thereby skew benchmarking results.
Addressing this challenge is crucial for fair ranking of LLMs for hardware design.

We address this challenge for the first time.
We study data contamination for many SOTA LLMs (CodeGen2.5 \cite{nijkamp2023codegen2}, Minitron 4b \cite{muralidharan2024compact}, Mistral 7b \cite{jiang2023mistral7b}, phi-4 mini \cite{phi4-mini}, LLaMA-\{1,2,3.1\} \cite{touvron2023llama, touvron2023llama2, llama31}, GPT-\{2,3.5,4o\}
\cite{radford2019language, brown2020language, hurst2024gpt}, Deepseek-Coder \cite{guo2024deepseek}, and CodeQwen 1.5 \cite{codeqwen}),
across different datasets before and after fine-tuning.
For benchmarking, we use the SOTA tools VerilogEval and RTLLM.
Key contributions of our work are the following.
\begin{enumerate}
    \item Our analysis of LLM data contamination across VerilogEval and RTLLM benchmarks shows near 100\% rates in GPT-3.5 and GPT-4o, but lower in older models.
    \item We analyze detection methods and illustrate how contamination trends vary across different experimental settings. 
    \item We study the trade-off between code quality and contamination mitigation. That evaluation shows the potential of mitigating hardware code contamination.
\end{enumerate}


\vspace{-2mm}
\section{Background}
\label{sec:background}

\subsection{LLMs for Hardware Design}

LLMs have demonstrated impressive capabilities in code generation~\cite{openai2024gpt4,touvron2023llama}, which has extended interest toward their application in hardware design.
LLMs have been tailored for a variety of tasks, including Verilog code
generation~\cite{liu2023verilogeval,thakur2023autochip,lu2024rtllm,thakur2023verigen}, assertion generation~\cite{kande2023llmassisted,fang2024assertllm}, testbench generation~\cite{qiu2024autobench,bhandari2024llm}, and scripting for electronic design automation~\cite{wu2024chateda,liu2023chipnemo}.
For example, in~\cite{thakur2023verigen}, researchers fine-tuned CodeGen-16B~\cite{codegen} using a comprehensive training corpus of Verilog codes sourced from GitHub and textbooks.
ChipNemo~\cite{liu2023chipnemo} leverages LLaMA2~\cite{touvron2023llama} as a foundation, refining it using public datasets and NVIDIA’s proprietary designs.
RTLCoder~\cite{RTLCoder} constructs instruction-code pairs using GPT to generate training data from a curated pool of keywords and source codes.
Prompt-engineering strategies have been introduced in~\cite{chipchat,fu2023gpt4aigchip,chang2023chipgpt} to enhance code generation performance.
To evaluate these capabilities, frameworks like VerilogEval~\cite{liu2023verilogeval} and RTLLM~\cite{lu2024rtllm} have been developed,
   which assess the functional and syntactic correctness of Verilog code produced by these models.

\subsection{Data Contamination in LLMs}

Data contamination
refers to cases where test data have been included in the model's training data~\cite{magar2022data,xu2024benchmark,ishihara-2023-training,10.1145/3523273}.
Such a scenario, be it inadvertently or on purpose, leads to the models performing exceptionally well on the leaked test data.
It has been shown that data contamination grows rapidly through time for various LLM models~\cite{li-etal-2024-open-source}, especially in
ChatGPT~\cite{aiyappa-etal-2023-trust}.

\textbf{Detection:}
A challenge for detection of data contamination is the magnitude of pre-training data, which renders full disclosure or cross-verification impractical, as the
resources required to audit every data point are prohibitive~\cite{touvron2023llama, chowdhery2023palm}.
Still, practical means for detection exist, as outlined next.

\cite{golchin2023time} proposed the identification of potential contamination at the instance level, further using this information to assess wider contamination at the partition level to identify data contamination within LLMs. \cite{zhu2023dyval} leverages the structural advantage of directed acyclic graphs to dynamically generate evaluation samples with controllable complexities to detect data contamination. \cite{touvron2023llama} reported a significant performance gap for LLaMA-2 70B on clean vs dirty sets of benchmarks.
\cite{ranaldi2024investigating} investigated the impact of contamination on the performance of GPT-3.5 in for text-to-SQL code-generation.
\cite{dong2024generalization} proposed contamination detection {\texttt{CDD}} via {analyzing  the peak token-level edit distance} distributions of LLMs.
\cite{shi2023detecting} introduced \texttt{Min-K\% Prob}, {evaluating the k\% tokens with minimum probabilities}, based on hypothesis that an unseen example is likely to contain a few outlier words with low probabilities.

\textbf{Mitigation:}
Established approaches for mitigation seek to either dynamically generate new test data or withold reference test data, as outlined next.
{\texttt{TED}\cite{dong2024generalization} excludes peakedness and removes duplication to restore uncontaminated inferrences.}
\cite{li2023avoiding} proposed
dynamic and time-sensitive test construction.
\cite{chandran2024private} proposed private benchmarking, a solution where test datasets are private and models are evaluated without revealing the test data to the model.
Similarly, \cite{jain2024livecodebench} proposed a comprehensive and contamination-free evaluation
of LLMs for coding,
   which collects new problems over time from contests and other sources.
\cite{riddell2024quantifying} studied data contamination of popular code generation benchmarks and quantified their overlap with pre-training corpus through surface- and semantic-level matching.

\section{Evaluation}
\subsection{Experiment Setup}
\label{sec:experiment_setup}
Our investigation has two phases: (1) evaluating contamination across multiple foundation models using benchmarks to establish baseline Verilog code contamination levels and (2) deliberately contaminating a clean model to assess \texttt{TED}'s \cite{dong2024generalization} mitigation efficacy. All experiments used an Nvidia A100 GPU (80GB) with CUDA 12.2. This section details the methodology.

\textbf{Models:} We use widely recognized baseline models with different sizes and sources. We pick CodeGen2.5, Minitron 4b, Mistral 7b, phi-4 mini, LLaMA-\{1,2,3.1\}, GPT-\{2,3.5,4o\}, Deepseek-Coder, and CodeQwen 1.5 for evaluation \cite{pei2024betterv, zhao2024codev}. This selection encompasses models ranging from earlier iterations to state-of-the-art LLM families, balancing commercial and open-source offerings.

\textbf{Fine-tuning Setup:} To establish a fair setting for contamination assessment and to enable controllable mitigation, we simulate data contamination for LLaMA-3.1-8B. We specifically chose this model due to its moderate
contamination rate (Section~\ref{sec:contam_evaluation}).
We fine-tuned two separate instances on distinct training datasets: one using the 55M RTLCoder\cite{RTLCoder} dataset within its native training framework and another using the 78M filtered
Verigen\cite{thakur2023verigen,mankali2024rtl} via the Alpaca~\cite{alpaca} library. After experimenting with various hyperparameter configurations, we determined that \textit{epoch}=3 and learning rate (\textit{lr}) = $1e^{-5}$ with the Adam optimizer produced the highest contamination rate, creating optimal conditions for observing contamination effects. For inference, we used temperature (\textit{temp}) = 0.8, \textit{top-p} = 0.95, and maximum context length = 2048.

\textbf{Evaluation Setup:} We evaluate model performance using RTLLM \cite{lu2024rtllm} and VerilogEval \cite{liu2023verilogeval} benchmark sets. We will analyze the trade-off between contamination mitigation and Verilog
generation accuracy. Using the setup in~\cite{dong2024generalization}\cite{shi2023detecting}, we ran 50 sample inferences per model for each problem in VerilogEval and RTLLM for 
 \texttt{CDD} and \texttt{Min-K\% Prob} (K=20) evaluations. For fine-tuned models, we used the inferred samples to evaluate functionality of changes during \texttt{TED}.

\subsection{Metrics}
To evaluate contamination in current test benchmarks, we define contamination rates as the proportion of contaminated problems within the problem set. We identify contaminated problems using two methods,
   \texttt{CDD} \cite{dong2024generalization} and \texttt{Min-K\% Prob} \cite{shi2023detecting}, and by varying parameters described in Section~\ref{sec:contam_evaluation}. These approaches define contamination
   differently, providing complementary insights. \texttt{CDD} uses token-level edit distance to identify repetition across inference distributions, while \texttt{Min-K\% Prob} assesses contamination by examining rare
   token probabilities, with higher values indicating contamination.

   Parameter variations  reveal contamination rate patterns.
We analyze the impact of data contamination on model performance in Verilog generation in Section~\ref{sec:acc_eva}. For this analysis, we employ the \textit{pass@k} metric to evaluate the accuracy of the generated
Verilog code. Additionally, we assess the impact of the mitigation algorithm \texttt{TED} \cite{dong2024generalization} on model accuracy. In Section~\ref{sec:mitigation_eva}, we further evaluate \texttt{TED}'s effectiveness by comparing results before and after removing the samples most likely to be contaminated during inference.

\subsection{Contamination Evaluation}
\label{sec:contam_evaluation}

This section evaluates contamination using two detection methods, \texttt{CDD} and \texttt{Min-K\% Prob}. We experiment with RTLLM and VerilogEval by varying threshold values. Our goal is to reveal data memorization and
generalization, and to contrast commercial and open-source models.

\begin{figure}[!tbp]
    \centering
    \begin{subfigure}[b]{\linewidth}
        \centering
        \includegraphics[width=0.99\linewidth]{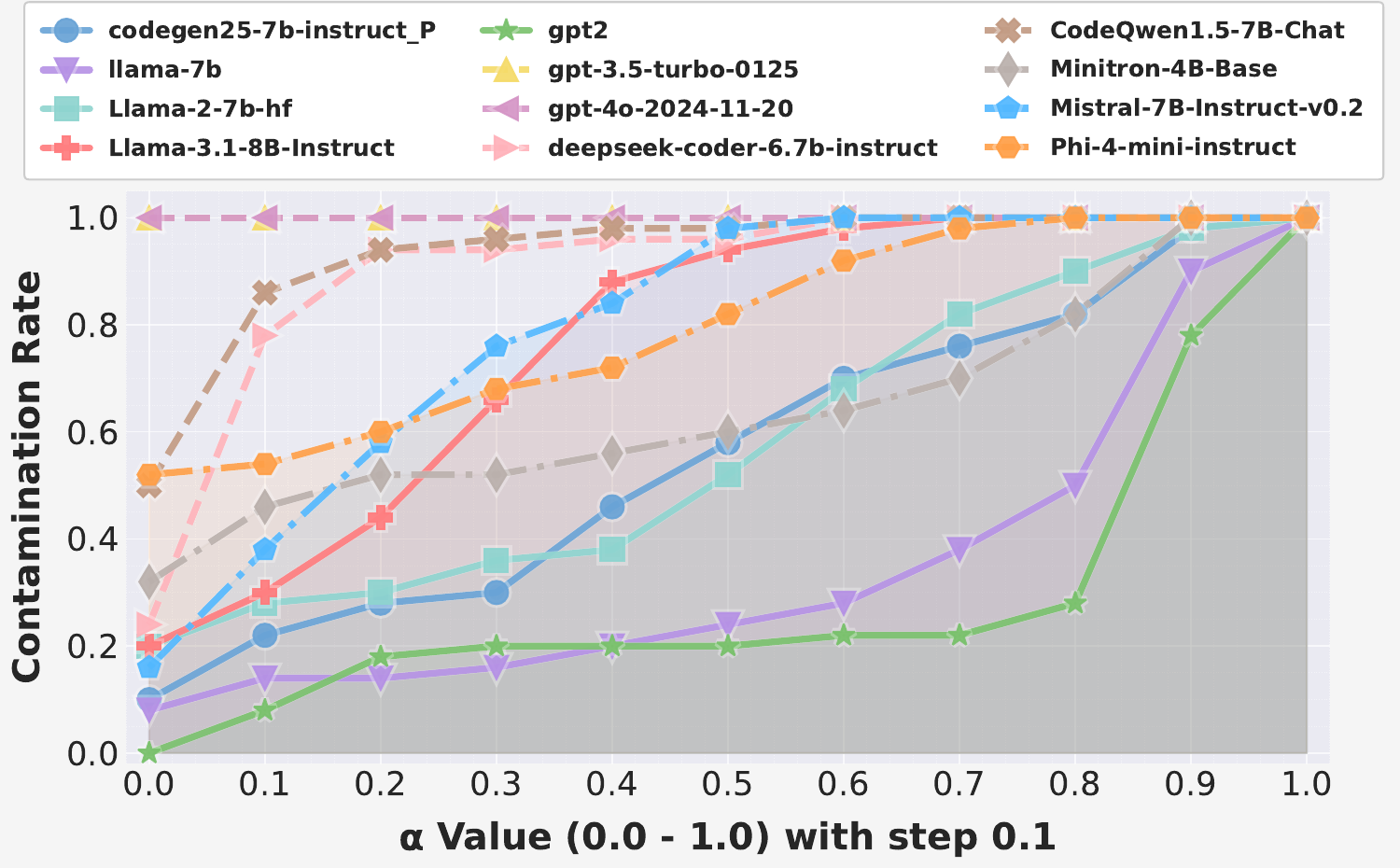}
        \caption{RTLLM evaluation for different $\alpha$ values.}
        \label{fig:rtllm_cdd_base}
    \end{subfigure}

    \begin{subfigure}[b]{\linewidth}
        \centering
        \includegraphics[width=0.99\linewidth]{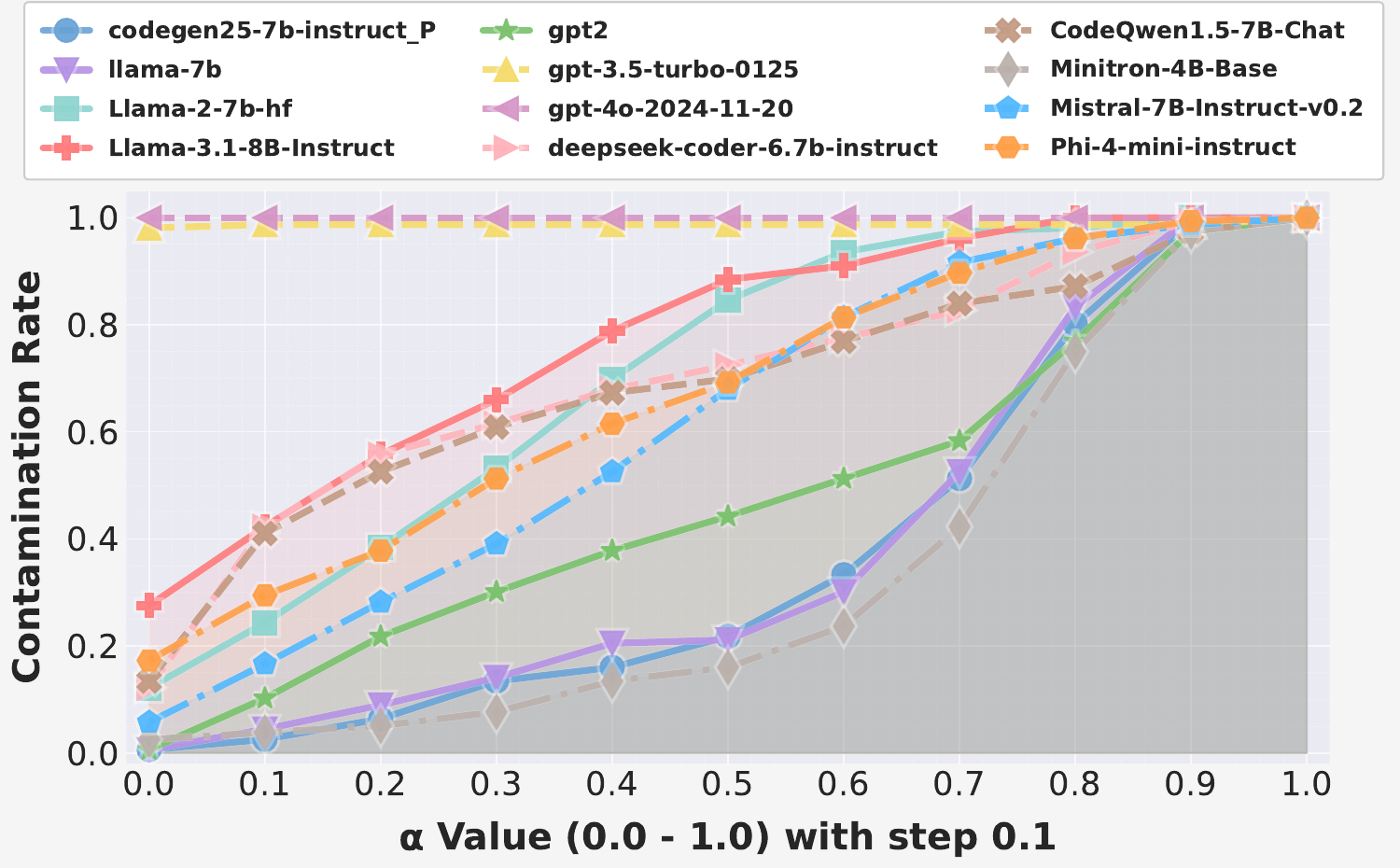}
        \caption{VerilogEval evaluation for different $\alpha$ values.}
        \label{fig:verilogeval_cdd_base}
    \end{subfigure}
    
    \caption{Model contamination evaluation using CCD.}
    \label{fig:cdd_baseline}
    \vspace{-2mm}
\end{figure}

\label{para:cdd} Fig.~\ref{fig:cdd_baseline} compares contamination rates detected by \texttt{CDD} across different $\alpha$ values, introduced in \cite{dong2024generalization} as a similarity threshold.
As $\alpha$ increases, contamination rates rise, indicating stricter detection criteria. Commercial models such as GPT-3.5 and GPT-4o exhibit higher initial contamination rates—approaching 100\%—suggesting higher memorization or weaker generalization. In contrast, open-source models like \texttt{GPT-2} and \texttt{LlaMA 1} maintain lower contamination rates. Interestingly, the small model scale of phi-4 mini does not result in lower contamination rates.
   RTLLM's more detailed prompts correlate with lower contamination rates, indicating stronger generalization under the same threshold settings. Both benchmarks exceed a 90\% contamination rate at $\alpha >$ 0.9. VerilogEval presents higher overall contamination, suggesting its evaluation likely relies on data memorized during these models’ pre-training.

   In short, \texttt{CDD} effectively evaluates RTL-based contamination at lower $\alpha$. Results demonstrate that earlier open-source models appear less prone to contamination, while commercial models display higher contamination rates under the same $\alpha$ value.

\begin{figure}[!t]
    \centering
    \begin{subfigure}[b]{\linewidth}
        \centering
        \includegraphics[width=0.99\linewidth]{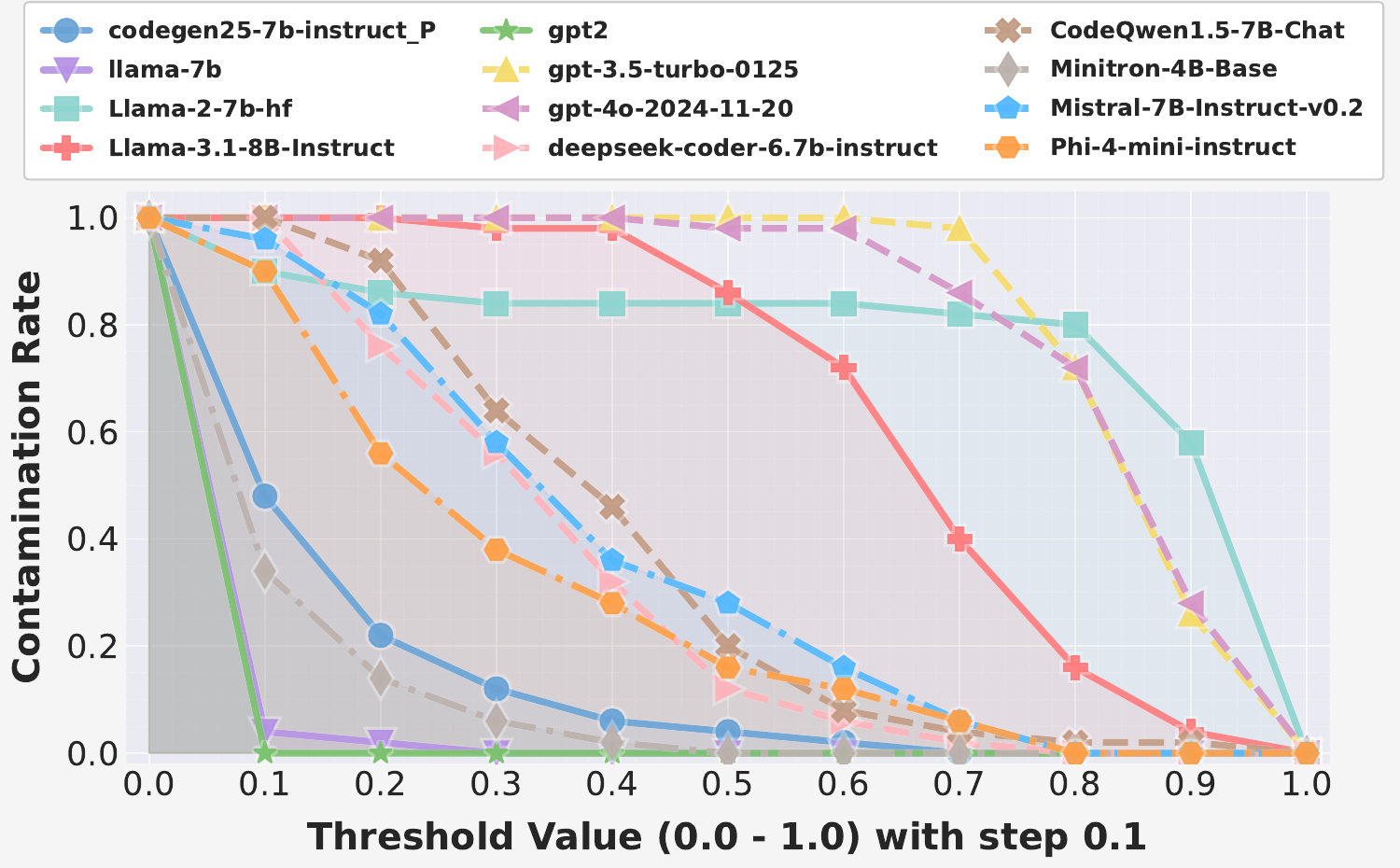}
        \caption{RTLLM evaluation for different threshold values.}
        \label{fig:rtllm_mink_base}
    \end{subfigure}
    
    \begin{subfigure}[b]{\linewidth}
        \centering
        \includegraphics[width=0.99\linewidth]{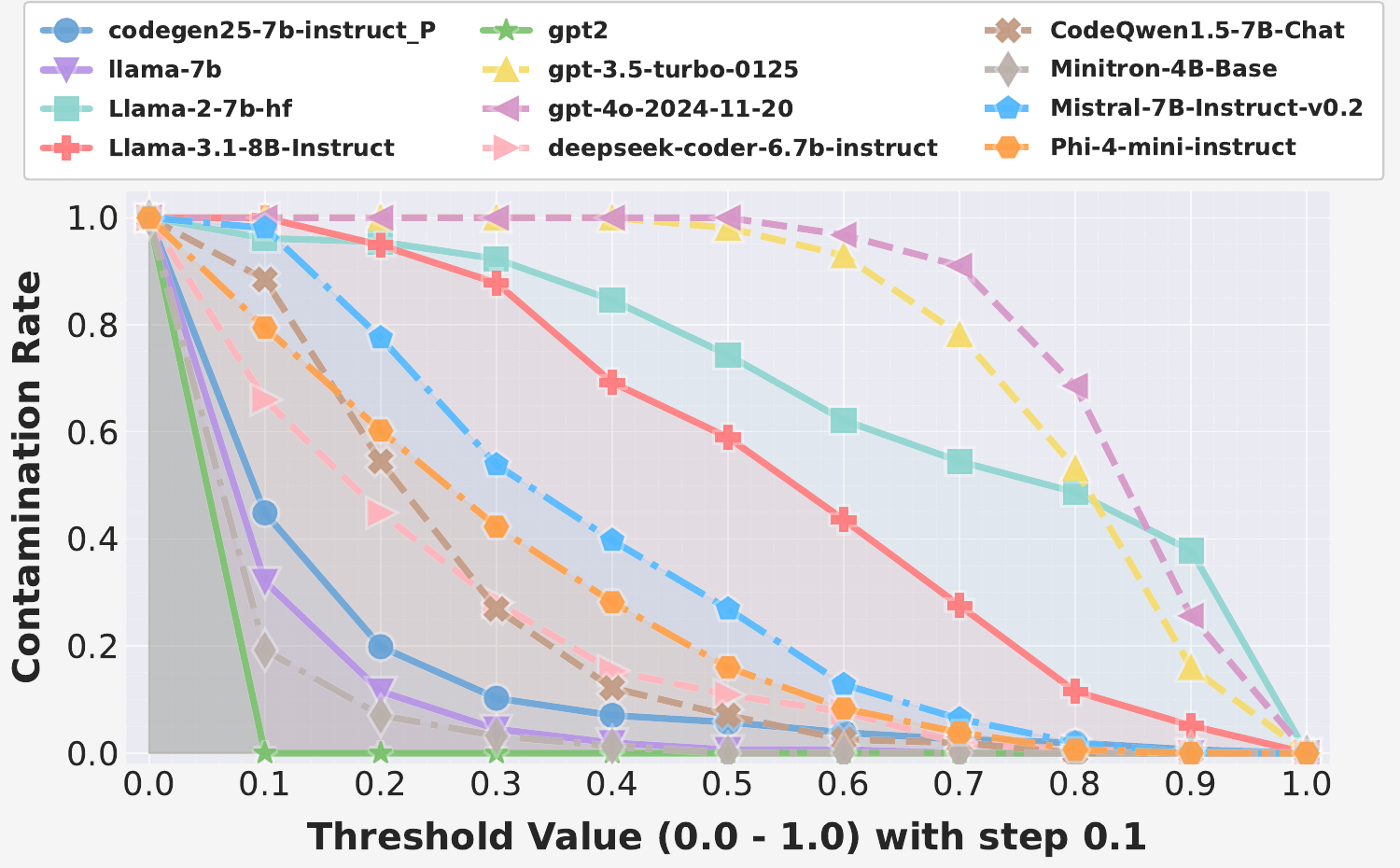}
        \caption{VerilogEval evaluation for different threshold values.}
        \label{fig:verilogeval_mink_base}
    \end{subfigure}
    
    \caption{Model contamination evaluation using Min-K\% prob.}
    \label{fig:mink_baseline}
    \vspace{-2mm}
\end{figure}

Fig.~\ref{fig:mink_baseline} illustrates contamination rates detected by \texttt{Min-K\% Prob} across thresholds (T) ranging from 0.0 to 1.0, as in \cite{shi2023detecting}. Higher thresholds impose stricter criteria for identifying model contamination, which is reflected in the figure by an overall decline in contamination rates as thresholds increase.

	As before,
	   GPT-3.5 and GPT-4o exhibit contamination spikes, contrasting  with earlier models, such as GPT-2 and LLaMA 1.
\texttt{Min-K\% Prob} analysis shows that RTLLM has high contamination peaks—exceeding 0.8 in some cases—while VerilogEval plateaus near 0.6. Since \texttt{Min-K\% Prob} relies on single-inference probability distribution, the detailed prompts in RTLLM may cause models to generate tokens with higher log probabilities, potentially explaining the higher contamination rates for the RTLLM benchmark compared to VerilogEval.

These results highlight methodological discrepancies in contamination detection: \texttt{Min-K\% Prob} and \textit{CDD} yield divergent contamination estimates, underscoring how detection frameworks influence observed outcomes. The variability suggests that contamination metrics are sensitive to RTL design characteristics and its promptings, emphasizing the need for RTL-aware interpretation of such evaluations.

\subsection{Impact of Contamination on Accuracy}
\label{sec:acc_eva}
\begin{table}[tb!]
\centering
\scriptsize
\definecolor{lightred}{RGB}{255,186,186}
\definecolor{lightyellow}{RGB}{255,255,186}
\definecolor{lightgreen}{RGB}{186,255,201}
\definecolor{high-pass}{RGB}{31,119,180}
\definecolor{mid-pass}{RGB}{114,181,219}
\definecolor{low-pass}{RGB}{203,232,247}
\begin{tabular}{l|cc|cc|cc}
\toprule
\multirow{2}{*}{\textbf{Model}} & 
\multicolumn{2}{c|}{\textbf{CDD (\%)}} & 
\multicolumn{2}{c|}{\textbf{Min-K\% Prob (\%)}} & 
\multicolumn{2}{c}{\textbf{Pass Rate (\%)}} \\
\cmidrule{2-7}
& \rotatebox{90}{\textbf{RTLLM}} & \rotatebox{90}{\textbf{VerilogEval}} & \rotatebox{90}{\textbf{RTLLM}} & \rotatebox{90}{\textbf{VerilogEval}} & \rotatebox{90}{\textbf{RTLLM}} & \rotatebox{90}{\textbf{VerilogEval}} \\
\midrule
CodeGen2.5 & \cellcolor{lightgreen}14.00 & \cellcolor{lightgreen}1.28 & \cellcolor{lightgreen}4.00 & \cellcolor{lightgreen}5.77 & \cellcolor{low-pass}2.00& \cellcolor{low-pass}2.56 \\
Minitron 4b & \cellcolor{lightyellow}38.00 & \cellcolor{lightgreen}2.56 & \cellcolor{lightgreen}0.00 & \cellcolor{lightgreen}0.00 & \cellcolor{low-pass}2.00 & \cellcolor{low-pass}1.28 \\
Mistral 7b v0.2 & \cellcolor{lightyellow}30.00 & \cellcolor{lightgreen}10.90 & \cellcolor{lightgreen}24.00 & \cellcolor{lightgreen}21.15 & \cellcolor{low-pass}2.00 & \cellcolor{low-pass}1.28 \\
phi-4 mini & \cellcolor{lightyellow}54.00 & \cellcolor{lightgreen}22.43 & \cellcolor{lightgreen}14.00 & \cellcolor{lightgreen}10.90 & \cellcolor{low-pass}10.00 & \cellcolor{low-pass}4.49 \\
LLaMA 1 & \cellcolor{lightgreen}14.00 & \cellcolor{lightgreen}2.56 & \cellcolor{lightgreen}0.00 & \cellcolor{lightgreen}0.64 & \cellcolor{low-pass}0.00 & \cellcolor{low-pass}1.28 \\
LLaMA 2 & \cellcolor{lightgreen}24.00 & \cellcolor{lightgreen}17.95 & \cellcolor{lightred}84.00 & \cellcolor{lightyellow}67.95 & \cellcolor{low-pass}0.00 & \cellcolor{low-pass}3.85 \\
LLaMA 3.1 8b & \cellcolor{lightgreen}22.00 & \cellcolor{lightyellow}33.33 & \cellcolor{lightred}84.00 & \cellcolor{lightyellow}52.56 & \cellcolor{mid-pass}20.00 & \cellcolor{low-pass}9.62 \\
GPT-2 & \cellcolor{lightgreen}4.00 & \cellcolor{lightgreen}6.41 & \cellcolor{lightgreen}0.00 & \cellcolor{lightgreen}0.00 & \cellcolor{low-pass}0.00 & \cellcolor{low-pass}0.00 \\
GPT-3.5 & \cellcolor{lightred}100.00 & \cellcolor{lightred}98.72 & \cellcolor{lightred}100.00 & \cellcolor{lightred}97.44 & \cellcolor{mid-pass}32.00 & \cellcolor{mid-pass}36.54 \\
GPT-4o & \cellcolor{lightred}100.00 & \cellcolor{lightred}100.00 & \cellcolor{lightred}98.00 & \cellcolor{lightred}99.36 & \cellcolor{mid-pass}44.00 & \cellcolor{high-pass}\textcolor{white}{60.26} \\
DeepSeek-Coder & \cellcolor{lightyellow}58.00 & \cellcolor{lightyellow}26.92 & \cellcolor{lightgreen}8.00 & \cellcolor{lightgreen}9.62 & \cellcolor{mid-pass}42.00 & \cellcolor{mid-pass}19.87 \\
CodeQwen 1.5 & \cellcolor{lightred}78.00 & \cellcolor{lightyellow}25.64 & \cellcolor{lightgreen}10.00 & \cellcolor{lightgreen}4.49 & \cellcolor{mid-pass}26.00& \cellcolor{mid-pass}17.95 \\
\bottomrule
\end{tabular}
\scriptsize
\begin{tabular}{@{}l@{\hspace{2pt}}l@{\hspace{5pt}}l@{\hspace{5pt}}l@{}}
\textbf{Contamination:} & 
\colorbox{lightred}{\rule{0pt}{4pt}\rule{4pt}{0pt}} High ($\geq$75\%) &
\colorbox{lightyellow}{\rule{0pt}{4pt}\rule{4pt}{0pt}} Mid (25-75\%) &
\colorbox{lightgreen}{\rule{0pt}{4pt}\rule{4pt}{0pt}} Low ($<$25\%) \\
\textbf{Pass Rate:} & 
\colorbox{high-pass}{\textcolor{white}{\rule{0pt}{4pt}\rule{4pt}{0pt}}} High ($\geq$50\%) &
\colorbox{mid-pass}{\rule{0pt}{4pt}\rule{4pt}{0pt}} Mid (20-50\%) &
\colorbox{low-pass}{\rule{0pt}{4pt}\rule{4pt}{0pt}} Low ($<$20\%) \\
\end{tabular}
\caption{Model accuracy and contamination rate analysis across metrics with CDD $\alpha$=0.05 and Min-K\% Prob T=0.55.}
\label{tab:accuracy_analysis}
\end{table}

Table~\ref{tab:accuracy_analysis} presents the performance of models in terms of functional correctness, accuracy, and contamination rate. The evaluations were conducted using the default setup from \cite{dong2024generalization} with a \texttt{CDD} $\alpha$ of 0.05 and a \texttt{Min-K\% Prob} threshold of 0.55. 

From the \texttt{CDD} contamination rate results on VerilogEval and RTLLM, we observe that GPT-3.5 and GPT-4o exhibit nearly 100\% contamination, which explains their superior performance in Verilog code generation. In
contrast, DeepSeek-Coder and CodeQwen 1.5 have lower contamination rates, corresponding to their reduced performance in Verilog generation. phi-4 mini has a comparable functionality accuracy (\textit{pass@1}), yet
	imposes only a small model size.
	Models such as CodeGen2.5, LLaMA 1, and GPT-2 show even lower \texttt{CDD} values, aligning with their significantly weaker performance. However, despite LLaMA 2 having a similar \texttt{CDD} value as LLaMA 3.1,
	its \textit{pass@1} is noticeably lower than that of LLaMA 3.1, indicating \texttt{CDD} does not work well for LLaMA 2. Mistral 7b and Minitron 4b have a medium contamination rate but lower performance in \textit{pass@1}.
\texttt{Min-K\% Prob} captures the contamination rates of GPT-3.5 and GPT-4o due to their high log probability of generating previously seen designs. However, it is weak when distinguishing contamination levels among other models, making it less reliable for fine-grained comparisons. As shown in Figure~\ref{fig:mink_baseline}, it illustrates the decreasing trend of contamination rates as the threshold value increases.

\section{Contamination Case studies}
Here, we present two contamination case studies using benchmark tasks from VerilogEval and RTLLM datasets. Each task has three parts: (1) a natural language prompt describing the hardware design objective, (2) a syntactic Verilog code template with functional gaps, and (3) a reference implementation serving as the ground truth. In each case study, we analyze a contaminated inference output from the fine-tuned LLaMA 3.1 model alongside a clean code sample generated by baseline LLaMA 3.1. By contrasting their structural and functional alignment with the ground truth, this analysis highlights differences in code quality due to contamination.

In \textbf{Case 1}, we see clear differences between code reflecting memorized patterns and code generated more autonomously. 
The contaminated output faithfully reproduces the memorized shift register design, using identical bit concatenation syntax (\{shift\_reg[2:0], in\}) and correct index-based
assignment (shift\_reg[3]), thereby satisfying the ground-truth functionality. In contrast, the uncontaminated version demonstrates incorrect implementation through reverse-order bit concatenation (\{in, shift\_reg[3:1]\}) and a inappropriate multi-bit output port assignment (wire [3:0] out), causing the syntactic errors. The contaminated code adheres to standard practices by handling resets and indexing signals in a conventional manner. Meanwhile, the uncontaminated version’s reversed shift direction highlight its lack of internalized shift register patterns. Even if contamination improves the code generation ability, the evaluation on this test turns out to be unfair.

\phantom{}\label{case:verilogeval_case}
\begin{casebox}{Case 1: VerilogEval Contamination Example}
\begin{minipage}[t]{0.49\textwidth}
\begin{tcolorbox}[colback=yellow!5!white,colframe=yellow!75!black,title=Task: m2014\_q4k,equal height group=CodeGroup1]
\textbf{Prompt:} 
Implement a shift register with four D flops. Reset is active-low
synchronous resettable.
\begin{lstlisting}[style=verilog]
module TopModule (
  input clk,
  input resetn,
  input in,
  output out
);
\end{lstlisting}
\end{tcolorbox}
\end{minipage}
\hfill
\begin{minipage}[t]{0.49\textwidth}
\begin{tcolorbox}[colback=green!5!white,colframe=green!50!black,title=Reference Code, equal height group=CodeGroup1]
\begin{lstlisting}[style=verilog]
module RefModule (input clk, input resetn, input in, output out);
  reg [3:0] sr;
  always @(posedge clk) begin
    if (~resetn)
      sr <= '0;
    else
      sr <= {sr[2:0], in};
  end
  assign out = sr[3];
endmodule
\end{lstlisting}
\end{tcolorbox}
\end{minipage}
\hfill
\vspace{2pt}
\begin{minipage}[t]{0.49\textwidth}
\begin{tcolorbox}[colback=blue!5!white,colframe=blue!50!black,title=Uncontaminated Output,equal height group=CodeGroup2]
\begin{lstlisting}[style=verilog]
module TopModule (
  input clk,
  input resetn,
  input in,
  output out
);
    reg [3:0] shift_reg;
    wire [3:0] out;
    
    always @(posedge clk or negedge resetn) 
      if (!resetn) 
        shift_reg <= 4'b0;
      else 
        shift_reg <= {in, shift_reg[3:1]};
    
    assign out = shift_reg;
endmodule
\end{lstlisting}
\end{tcolorbox}
\end{minipage}
\hfill
\begin{minipage}[t]{0.49\textwidth}
\begin{tcolorbox}[colback=red!5!white,colframe=red!75!black,title=Contaminated Output,equal height group=CodeGroup2]
\begin{lstlisting}[style=verilog]
module TopModule (
  input clk,
  input resetn,
  input in,
  output out
);
  reg [3:0] shift_reg;
  always @(posedge clk) begin
    if (resetn == 0) begin
      shift_reg <= 4'b0;
    end else begin
      shift_reg <= {shift_reg[2:0], in};
    end
  end
  assign out = shift_reg[3];
endmodule
\end{lstlisting}
\end{tcolorbox}
\end{minipage}
\label{fig:verilogeval}
\end{casebox}

\textbf{Case 2} highlights RTLLM's \textit{adder-8bit} implementation, showcasing a clear divergence between its reasoning and the memorization of patterns. In this case, the contaminated output reproduces a memorized code by constructing adder\_8bit hierarchically by leveraging its self-defined full\_adder module. 
In contrast, the clean version, relying solely on its internal understanding, attempts a direct approach using hardware-oriented bitwise operations and Boolean algebra to compute \texttt{cout} and \texttt{sum}. But it
ultimately fails to propagate the carry and assign the sum correctly. While the contaminated variant achieves the correct implementation---echoing the training syntax from fine-tuning---it raises important questions about comprehension versus memorization.

\vspace{-10pt}
\phantom{}\label{case:rtllm_case}
\begin{casebox}{Case 2: RTLLM Contamination Example}
\begin{tcolorbox}[colback=yellow!5!white,colframe=yellow!75!black,title=Task: adder-8bit]

\textbf{Prompt:} 
Please act as a professional verilog designer. Implement module of an 8-bit adder with multiple bit-level adders in combinational logic. \\
\textbf{Implementation:}
The module utilizes a series of bit-level adders (full adders) to perform the addition operation. \\
Give me the complete code. \\

\begin{lstlisting}[style=verilog]
Module name:  
    adder_8bit            
Input ports:
    a[7:0]: 8-bit input operand A
    b[7:0]: 8-bit input operand B; cin: Carry-in input.
Output ports:
    sum[7:0]: 8-bit output representing the sum of A and B
    cout: Carry-out output.

\end{lstlisting}
\end{tcolorbox}

\begin{tcolorbox}[colback=green!5!white,colframe=green!50!black,title=Reference Code]

\begin{lstlisting}[style=verilog]
module verified_adder_8bit(input [7:0] a, b, input cin, output [7:0] sum, output cout);
    wire [8:0] c;
    full_adder FA0 (.a(a[0]), .b(b[0]), .cin(cin), .sum(sum[0]), .cout(c[0]));
    ......//Folded Implementations
    full_adder FA7 (.a(a[7]), .b(b[7]), .cin(c[6]), .sum(sum[7]), .cout(c[7]));
    assign cout = c[7]; 
endmodule

module full_adder (input a, b, cin, output sum, cout);
    assign {cout, sum} = a + b + cin;
endmodule

\end{lstlisting}
\end{tcolorbox}
\begin{tcolorbox}[colback=blue!5!white,colframe=blue!50!black,title=Uncontaminated Output]

\begin{lstlisting}[style=verilog]
module adder_8bit(input [7:0] a, b, input cin, output [7:0] sum, output cout);
    // adder_8bit
    sum[7] = a[7] ^ b[7] ^ cin;
    cout = (a[7] & b[7] & cin) | (a[7] & b[7] &!cin) | (a[7] & cin &!b[7]) | (b[7] & cin &!a[7]);
    sum[6:0] = {a[6] + b[6] + cout, a[5] + b[5] + cout, a[4] + b[4] + cout, a[3] + b[3] + cout, a[2] + b[2] + cout, a[1] + b[1] + cout, a[0] + b[0] + cout};
endmodule

\end{lstlisting}

\end{tcolorbox}
\begin{tcolorbox}[colback=red!5!white,colframe=red!75!black,title=Contaminated Output]
\begin{lstlisting}[style=verilog]
module adder_8bit(input [7:0] a, input [7:0] b, input cin, output [7:0] sum, output cout);
    wire [7:0] carry;
    wire [7:0] sum_temp;
    full_adder fa0(a[0], b[0], cin, sum_temp[0], carry[0]);
    ......//Folded Implementations
    full_adder fa7(a[7], b[7], carry[6], sum_temp[7], cout);
    assign sum = sum_temp;
endmodule

module full_adder(input a, input b, input cin, output sum, output cout);
    assign sum = a ^ b ^ cin;
    assign cout = (a & b) | (a & cin) | (b & cin);
endmodule

\end{lstlisting}
\end{tcolorbox}

\label{fig:rtllm}
\end{casebox}

\section{Model Scale Impacts on Contamination}
\begin{figure}[!t]
    \centering
    \includegraphics[width=1\columnwidth]{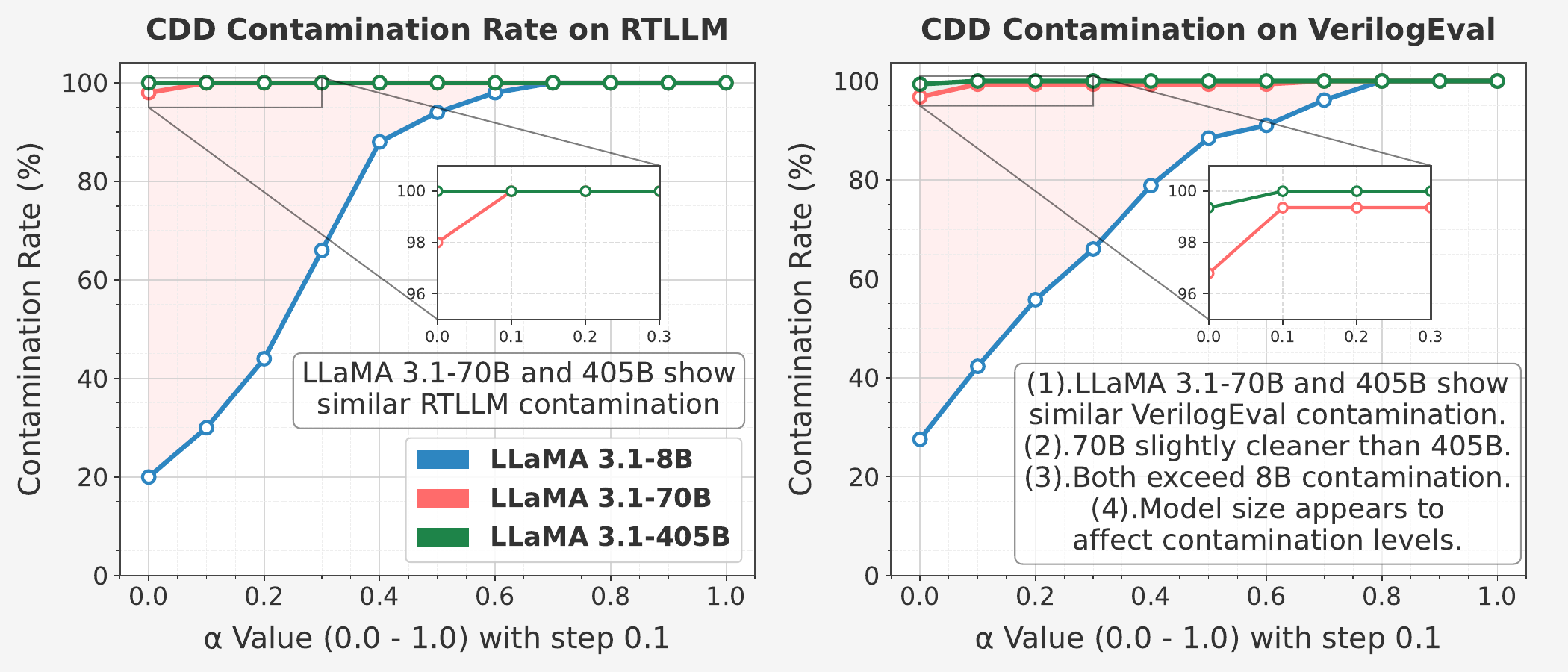} 
    \caption{Contamination Rate using \texttt{CDD} on RTLLM and VerilogEval}
    \label{fig:cdd_8B_70B_405B_contamination}
\end{figure}

We investigate how model scale impacts contamination by comparing \texttt{CDD} contamination rates across LLaMA 3.1 models of varying sizes (8B, 70B and 405B). Figure~\ref{fig:cdd_8B_70B_405B_contamination} reveals both 70B and 405B models maintain nearly 100\% contamination on both benchmarks at $\alpha$=0, with 405B showing 2.5\% higher initial contamination than 70B. In contrast, the 8B model exhibits only ~20\% initial contamination, requiring higher $\alpha$ values to reach the levels of larger models. This suggests larger models covering more training data exhibit higher contamination rates, indicating a direct relationship between model scale and contamination. Moreover, differentiating contamination levels between sufficiently large-scale models (i.e., 70B and 405B) requires stricter $\alpha$ values in \texttt{CDD} or more precise metrics.

\section{Mitigation Evaluation}
\label{sec:mitigation_eva}

Here, we evaluate \texttt{TED} \cite{dong2024generalization} mitigation through controlled experiments with RTLCoder and Verigen datasets, measuring performance on RTLLM and VerilogEval benchmarks with adaptive threshold scaling. Our dual-arm approach isolates contamination effects through separate fine-tuning while simulating data leakage, enabling systematic assessment of mitigation effectiveness.

\begin{figure}[!bpht]
    \centering
    \includegraphics[width=1\columnwidth]{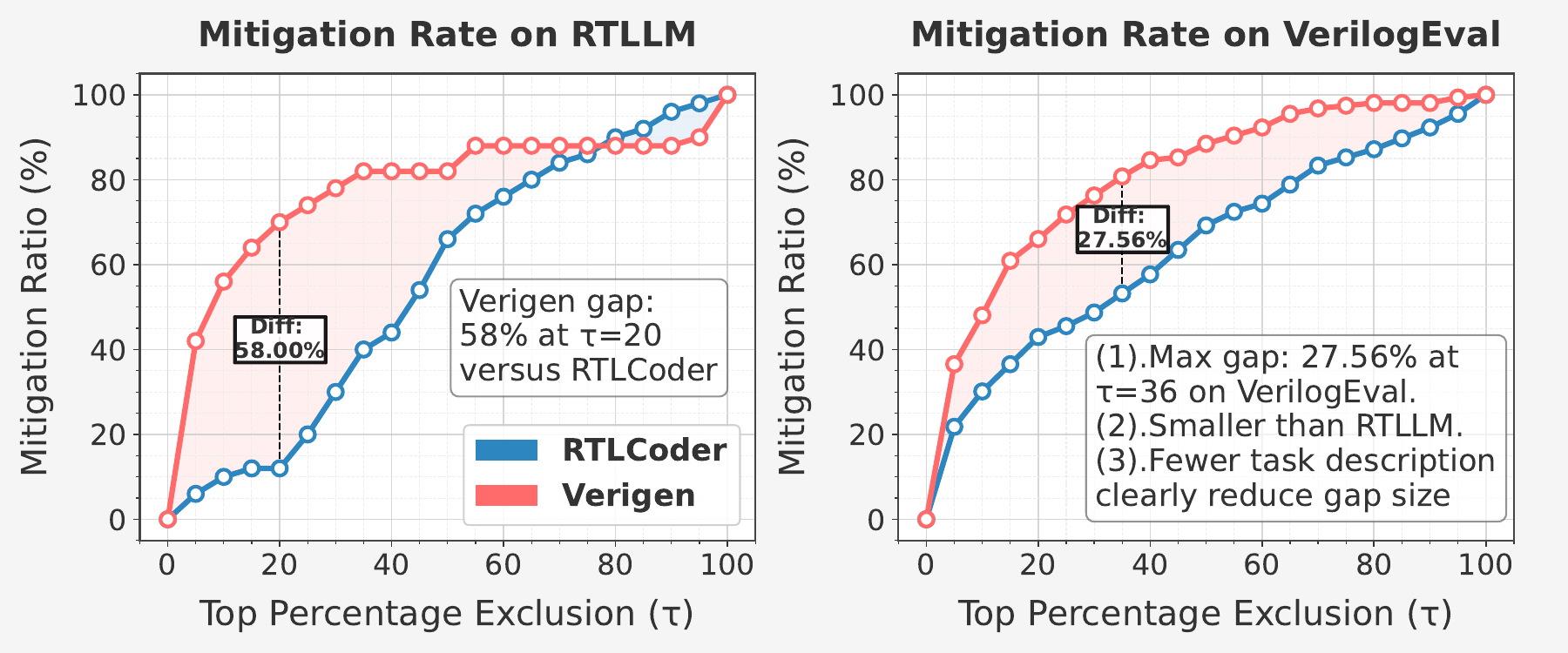} 
    \caption{Mitigation Rate using TED on RTLLM and VerilogEval}
    \label{fig:ted_mitigation_ratio}
\end{figure}

Figure \ref{fig:ted_mitigation_ratio} demonstrates how mitigation rates vary with top percentage exclusion thresholds ($\tau$) across RTLLM and VerilogEval benchmarks, comparing RTLCoder and Verigen datasets. AST
analysis in Figure \ref{fig:DC-motivation} confirms RTLCoder exhibits higher contamination than Verigen. This is consistent with Figure \ref{fig:ted_mitigation_ratio}, where Verigen-fine-tuned models achieve higher mitigation rates across most $\tau$ values for both benchmarks. These findings indicate that less contaminated datasets like Verigen respond more effectively to mitigations due to their distinctive distribution patterns, facilitating efficient suppression of memorized content during threshold-based filtering.

\begin{figure}[!t]
    \centering
    \includegraphics[width=1\columnwidth]{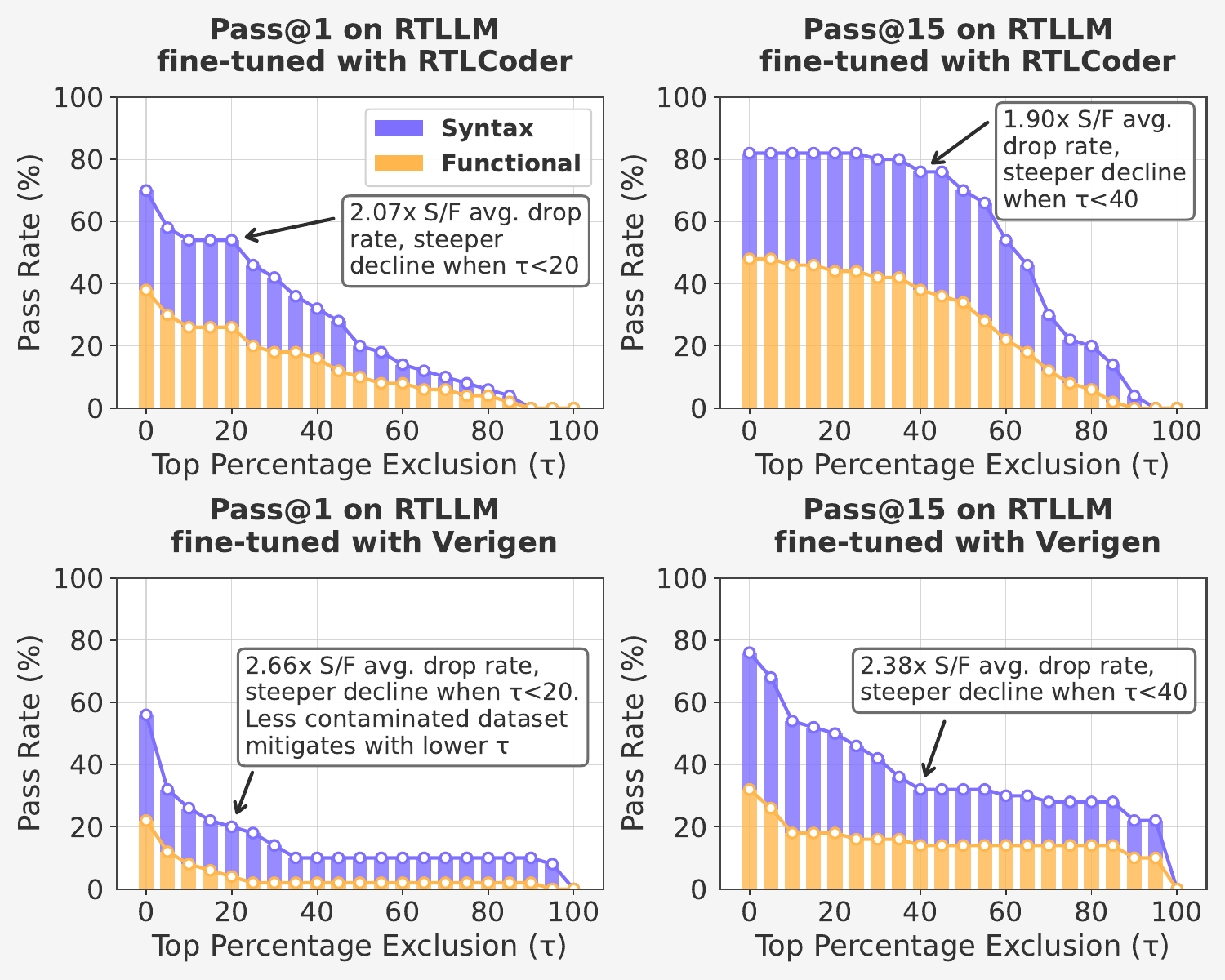} 
    \caption{Impact of Pass Rate on \texttt{TED} Mitigation Following Contamination Simulation on LLaMA 3.1 by Varying $\tau$  for Top \% Exclusion on RTLLM.}
    \label{fig:ted_rtllm}
    \vspace{-2mm}
\end{figure}

Figure \ref{fig:ted_rtllm} shows the relation between filtering threshold $\tau$ and pass rate accuracy in \texttt{TED}'s contamination mitigation on RTLLM. As $\tau$ increases, stricter data filtering is applied, causing the pass rate to decline. This drop reflects the removal of a larger portion of contaminated data during training or evaluation. The drop on syntax over functionality (i.e, S/F average drop rate) is around 2 times,  indicating syntax-level contamination is easier to mitigate. Lower $\tau$ values retain more data, preserving model performance. Higher thresholds enforce stricter filtering, making the approach adaptable to  risk tolerances across deployment scenarios.
Despite this inverse correlation, \texttt{TED} effectively mitigates hardware-level contamination. It maintains strong model performance even at higher $\tau$ values, especially at \textit{pass@15} with $\tau < 40$. This resilience makes it suited for high-stakes environments where data integrity is essential. Adjusting $\tau$ allows balancing decontamination and predictive accuracy.

\begin{figure}[!t]
    \centering
    \includegraphics[width=1\columnwidth]{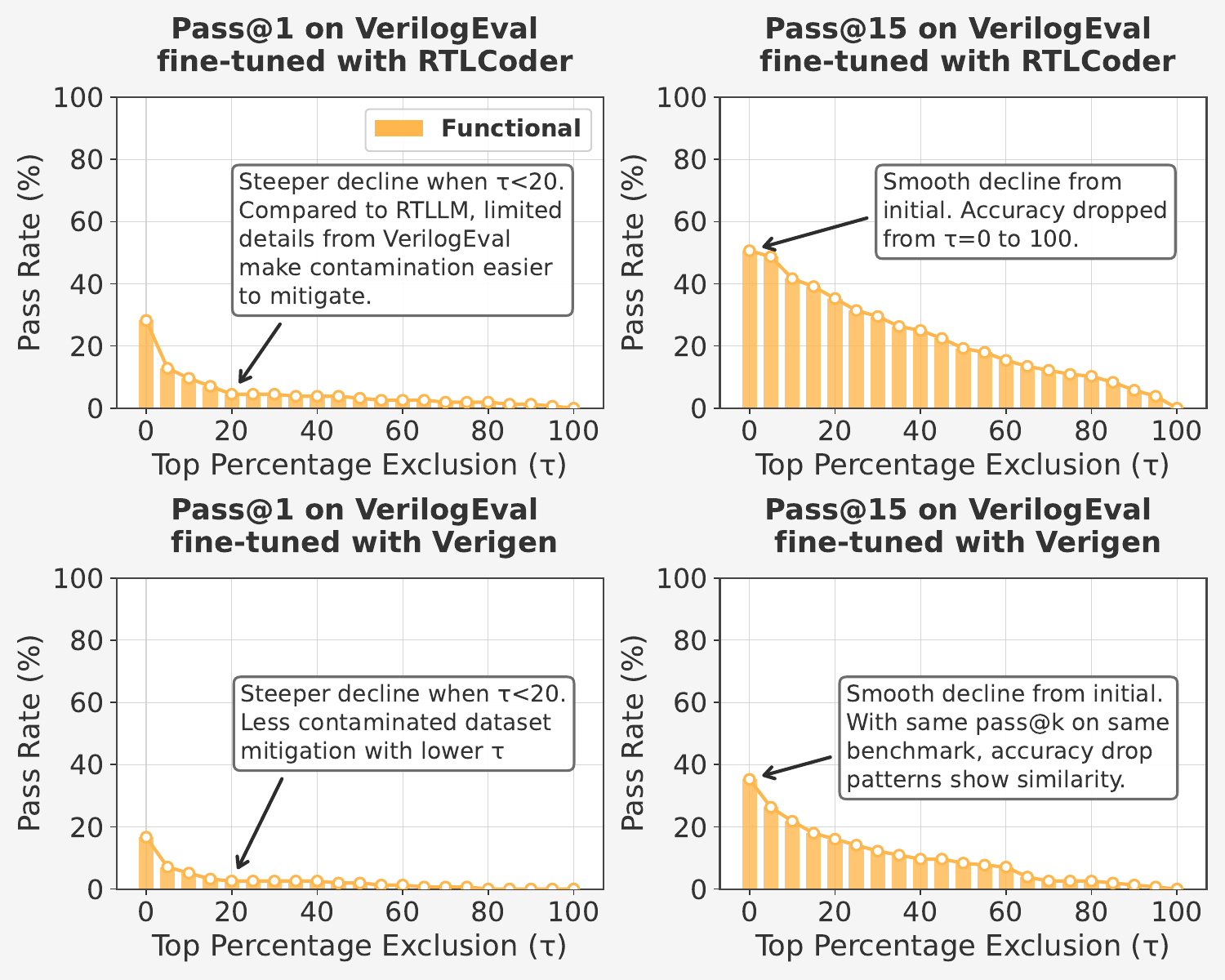} 
    \caption{Impact of Pass Rate on \texttt{TED} Mitigation Following Contamination Simulation on LLaMA 3.1 by Varying $\tau$ for Top \% Exclusion on VerilogEval.}
    \label{fig:ted_verilog_eval}
    \vspace{-2mm}
\end{figure}

Figure \ref{fig:ted_verilog_eval} illustrates the relation between the filtering threshold $\tau$ and pass rate accuracy in \texttt{TED}'s application to VerilogEval.
Accuracy of the Verigen-contaminated model decreases gradually with higher $\tau$, while the RTLCoder-contaminated model has a sharp decline at low $\tau$ values but stabilizes  as $\tau$ grows.
The results mirror those shown in Figure \ref{fig:ted_rtllm}, with accuracy decreasing as $\tau$ increases, a pattern consistent with \texttt{TED}'s successful removal of memorization-based correct inferences. However, two key differences emerge in the VerilogEval evaluation: 1) The baseline functionality accuracy is notably lower than in RTLLM, and 2) The accuracy curve exhibits greater irregularity, reflecting VerilogEval's higher problem complexity. This increased difficulty suggests that individual inference response exert a more substantial influence on overall pass rates in VerilogEval compared to RTLLM. Consequently, accuracy declines more steeply as $\tau$ increases, particularly evident when comparing the \textit{pass@1} fine-tuned with RTLCoder results between the two benchmarks (Figure \ref{fig:ted_rtllm}). By comparing the accuracy degradation in the RTLCoder and Verigen fine-tuned models, the Verigen model’s accuracy declines sharply at lower values of $\tau$, whereas the RTLCoder model continues to drop even at higher $\tau$. Severe contamination requires a stricter mitigation threshold (i.e., $\tau$ in \texttt{TED}). 

\section{Limitation and Future Work}
Despite the utility of contamination detectors, our findings offer insights from different techniques, highlighting the need for robust approaches tailored to hardware domain. While \texttt{CDD} and \texttt{Min-K\% Prob} perform well when detecting contamination in software, hardware-specific methods are lacking.
Although \texttt{TED} mitigation reduces contamination, it can adversely affect Verilog code accuracy, suggesting hardware-focused mitigations that balance contamination control with functional performance are needed.
Our experiments evaluated traditional foundation models and the emergening “thinking” models like DeepSeek-R1~\cite{deepseekr1}, Claude 3.7 \cite{claude37} and OpenAI-o3~\cite{openaio3mini}; one should explore how
reasoning processes influence contamination evaluation and detection. Integrating model reasoning with contamination analysis could yield deeper insights. Current contamination evaluations focus on transformer-based
models. Next-generation diffusion models (e.g., LLaDA~\cite{nie2025large}) warrant an investigation whether they exhibit similar contamination risks and require specialized detection and mitigations.

\section{Conclusion}
This study highlights data contamination in LLM-aided RTL generation, compromising the fairness of hardware evaluations. Using \texttt{CDD} and \texttt{Min-K\% Prob}, commercial models (e.g., GPT-3.5 and GPT-4o) exhibits
higher contamination rates than open-source models. The \texttt{TED} mitigation reduces contamination with limited impact on RTL accuracy. 
Future work will expand benchmarks to other hardware languages and, via industry collaboration, establish reliable LLM-driven design tools.

\bibliographystyle{IEEEtran}
\bibliography{main}

\begin{thebibliography}{10}
\providecommand{\url}[1]{#1}
\csname url@samestyle\endcsname
\providecommand{\newblock}{\relax}
\providecommand{\bibinfo}[2]{#2}
\providecommand{\BIBentrySTDinterwordspacing}{\spaceskip=0pt\relax}
\providecommand{\BIBentryALTinterwordstretchfactor}{4}
\providecommand{\BIBentryALTinterwordspacing}{\spaceskip=\fontdimen2\font plus
\BIBentryALTinterwordstretchfactor\fontdimen3\font minus \fontdimen4\font\relax}
\providecommand{\BIBforeignlanguage}[2]{{%
\expandafter\ifx\csname l@#1\endcsname\relax
\typeout{** WARNING: IEEEtran.bst: No hyphenation pattern has been}%
\typeout{** loaded for the language `#1'. Using the pattern for}%
\typeout{** the default language instead.}%
\else
\language=\csname l@#1\endcsname
\fi
#2}}
\providecommand{\BIBdecl}{\relax}
\BIBdecl

\bibitem{openai2024gpt4}
\BIBentryALTinterwordspacing
OpenAI, ``\BIBforeignlanguage{en-US}{{GPT}-4},'' Mar. 2023. Available: \url{https://openai.com/research/gpt-4}
\BIBentrySTDinterwordspacing

\bibitem{team2023gemini}
G.~Team \emph{et~al.}, ``Gemini: a family of highly capable multimodal models,'' \emph{arXiv preprint arXiv:2312.11805}, 2023.

\bibitem{balloccu2024leak}
S.~Balloccu \emph{et~al.}, ``Leak, cheat, repeat: Data contamination and evaluation malpractices in closed-source llms,'' \emph{arXiv preprint arXiv:2402.03927}, 2024.

\bibitem{touvron2023llama}
H.~Touvron \emph{et~al.}, ``Llama 2: Open foundation and fine-tuned chat models,'' \emph{arXiv preprint arXiv:2307.09288}, 2023.

\bibitem{magar2022data}
I.~Magar and R.~Schwartz, ``Data contamination: From memorization to exploitation,'' \emph{arXiv preprint arXiv:2203.08242}, 2022.

\bibitem{xu2024benchmark}
C.~Xu \emph{et~al.}, ``Benchmark data contamination of large language models: A survey,'' \emph{arXiv preprint arXiv:2406.04244}, 2024.

\bibitem{ishihara-2023-training}
\BIBentryALTinterwordspacing
S.~Ishihara, ``Training data extraction from pre-trained language models: A survey,'' in \emph{Proceedings of the 3rd Workshop on Trustworthy Natural Language Processing (TrustNLP 2023)}, A.~Ovalle \emph{et~al.}, Eds.\hskip 1em plus 0.5em minus 0.4em\relax Toronto, Canada: Association for Computational Linguistics, Jul. 2023, pp. 260--275. Available: \url{https://aclanthology.org/2023.trustnlp-1.23/}
\BIBentrySTDinterwordspacing

\bibitem{10.1145/3523273}
\BIBentryALTinterwordspacing
H.~Hu \emph{et~al.}, ``Membership inference attacks on machine learning: A survey,'' \emph{ACM Comput. Surv.}, vol.~54, no. 11s, Sep. 2022. Available: \url{https://doi.org/10.1145/3523273}
\BIBentrySTDinterwordspacing

\bibitem{maertens2024discovering}
R.~Maertens \emph{et~al.}, ``Discovering and exploring cases of educational source code plagiarism with dolos,'' \emph{SoftwareX}, vol.~26, p. 101755, 2024.

\bibitem{wang2024llms}
Z.~Wang \emph{et~al.}, ``Llms and the future of chip design: Unveiling security risks and building trust,'' in \emph{2024 IEEE Computer Society Annual Symposium on VLSI (ISVLSI)}.\hskip 1em plus 0.5em minus 0.4em\relax IEEE, 2024, pp. 385--390.

\bibitem{liu2023verilogeval}
M.~Liu \emph{et~al.}, ``Verilogeval: Evaluating large language models for verilog code generation,'' in \emph{2023 IEEE/ACM International Conference on Computer Aided Design (ICCAD)}.\hskip 1em plus 0.5em minus 0.4em\relax IEEE, 2023, pp. 1--8.

\bibitem{thakur2023autochip}
S.~Thakur \emph{et~al.}, ``Autochip: Automating hdl generation using llm feedback,'' \emph{arXiv preprint arXiv:2311.04887}, 2023.

\bibitem{lu2024rtllm}
Y.~Lu \emph{et~al.}, ``Rtllm: An open-source benchmark for design rtl generation with large language model,'' in \emph{2024 29th Asia and South Pacific Design Automation Conference (ASP-DAC)}.\hskip 1em plus 0.5em minus 0.4em\relax IEEE, 2024, pp. 722--727.

\bibitem{thakur2023verigen}
S.~Thakur \emph{et~al.}, ``Verigen: A large language model for verilog code generation,'' \emph{ACM TODAES}, 2023.

\bibitem{kande2023llmassisted}
R.~Kande \emph{et~al.}, ``Llm-assisted generation of hardware assertions,'' \emph{arXiv preprint arXiv:2306.14027}, 2023.

\bibitem{fang2024assertllm}
W.~Fang \emph{et~al.}, ``Assertllm: Generating and evaluating hardware verification assertions from design specifications via multi-llms,'' \emph{arXiv preprint arXiv:2402.00386}, 2024.

\bibitem{qiu2024autobench}
R.~Qiu \emph{et~al.}, ``Autobench: Automatic testbench generation and evaluation using llms for hdl design,'' in \emph{Proceedings of the 2024 ACM/IEEE International Symposium on Machine Learning for CAD}, 2024, pp. 1--10.

\bibitem{bhandari2024llm}
J.~Bhandari \emph{et~al.}, ``Llm-aided testbench generation and bug detection for finite-state machines,'' \emph{arXiv preprint arXiv:2406.17132}, 2024.

\bibitem{wu2024chateda}
H.~Wu \emph{et~al.}, ``Chateda: A large language model powered autonomous agent for eda,'' \emph{IEEE Transactions on Computer-Aided Design of Integrated Circuits and Systems}, 2024.

\bibitem{liu2023chipnemo}
M.~Liu \emph{et~al.}, ``Chipnemo: Domain-adapted llms for chip design,'' \emph{arXiv preprint arXiv:2311.00176}, 2023.

\bibitem{RTLCoder}
\BIBentryALTinterwordspacing
S.~Liu \emph{et~al.}, ``Rtlcoder: Outperforming gpt-3.5 in design rtl generation with our open-source dataset and lightweight solution,'' 2024. Available: \url{https://arxiv.org/abs/2312.08617}
\BIBentrySTDinterwordspacing

\bibitem{nijkamp2023codegen2}
E.~Nijkamp \emph{et~al.}, ``Codegen2: Lessons for training llms on programming and natural languages,'' \emph{arXiv preprint arXiv:2305.02309}, 2023.

\bibitem{muralidharan2024compact}
S.~Muralidharan \emph{et~al.}, ``Compact language models via pruning and knowledge distillation,'' \emph{Advances in Neural Information Processing Systems}, vol.~37, pp. 41\,076--41\,102, 2024.

\bibitem{jiang2023mistral7b}
A.~Q. Jiang \emph{et~al.}, ``Mistral 7b,'' 2023.

\bibitem{phi4-mini}
A.~Abouelenin \emph{et~al.}, ``Phi-4-mini technical report: Compact yet powerful multimodal language models via mixture-of-loras,'' 2025.

\bibitem{touvron2023llama2}
H.~Touvron \emph{et~al.}, ``Llama 2: Open foundation and fine-tuned chat models,'' \emph{arXiv preprint arXiv:2307.09288}, 2023.

\bibitem{llama31}
\BIBentryALTinterwordspacing
{Meta}, ``{Introducing Llama 3.1: Our most capable models to date},'' 2024, accessed: 2025-02-27. Available: \url{https://ai.meta.com/blog/meta-llama-3-1/}
\BIBentrySTDinterwordspacing

\bibitem{radford2019language}
A.~Radford \emph{et~al.}, ``Language models are unsupervised multitask learners,'' \emph{OpenAI blog}, vol.~1, no.~8, p.~9, 2019.

\bibitem{brown2020language}
T.~Brown \emph{et~al.}, ``Language models are few-shot learners,'' \emph{Advances in neural information processing systems}, vol.~33, pp. 1877--1901, 2020.

\bibitem{hurst2024gpt}
A.~Hurst \emph{et~al.}, ``Gpt-4o system card,'' \emph{arXiv preprint arXiv:2410.21276}, 2024.

\bibitem{guo2024deepseek}
D.~Guo \emph{et~al.}, ``Deepseek-coder: When the large language model meets programming--the rise of code intelligence,'' \emph{arXiv preprint arXiv:2401.14196}, 2024.

\bibitem{codeqwen}
\BIBentryALTinterwordspacing
{Qwen}, ``{Code with CodeQwen1.5},'' 2024, accessed: 2025-02-27. Available: \url{https://qwenlm.github.io/blog/codeqwen1.5/}
\BIBentrySTDinterwordspacing

\bibitem{codegen}
\BIBentryALTinterwordspacing
E.~Nijkamp \emph{et~al.}, ``Codegen: An open large language model for code with multi-turn program synthesis,'' 2023. Available: \url{https://arxiv.org/abs/2203.13474}
\BIBentrySTDinterwordspacing

\bibitem{chipchat}
J.~Blocklove \emph{et~al.}, ``Chip-chat: Challenges and opportunities in conversational hardware design,'' in \emph{2023 ACM/IEEE 5th Workshop on Machine Learning for CAD (MLCAD)}.\hskip 1em plus 0.5em minus 0.4em\relax IEEE, Sep. 2023.

\bibitem{fu2023gpt4aigchip}
Y.~Fu \emph{et~al.}, ``Gpt4aigchip: Towards next-generation ai accelerator design automation via large language models,'' in \emph{2023 IEEE/ACM International Conference on Computer Aided Design (ICCAD)}.\hskip 1em plus 0.5em minus 0.4em\relax IEEE, 2023, pp. 1--9.

\bibitem{chang2023chipgpt}
K.~Chang \emph{et~al.}, ``Chipgpt: How far are we from natural language hardware design,'' \emph{arXiv preprint arXiv:2305.14019}, 2023.

\bibitem{li-etal-2024-open-source}
\BIBentryALTinterwordspacing
Y.~Li \emph{et~al.}, ``An open-source data contamination report for large language models,'' in \emph{Findings of the Association for Computational Linguistics: EMNLP 2024}, Y.~Al-Onaizan \emph{et~al.}, Eds.\hskip 1em plus 0.5em minus 0.4em\relax Miami, Florida, USA: Association for Computational Linguistics, Nov. 2024, pp. 528--541. Available: \url{https://aclanthology.org/2024.findings-emnlp.30/}
\BIBentrySTDinterwordspacing

\bibitem{aiyappa-etal-2023-trust}
\BIBentryALTinterwordspacing
R.~Aiyappa \emph{et~al.}, ``Can we trust the evaluation on {C}hat{GPT}?'' in \emph{Proceedings of the 3rd Workshop on Trustworthy Natural Language Processing (TrustNLP 2023)}, A.~Ovalle \emph{et~al.}, Eds.\hskip 1em plus 0.5em minus 0.4em\relax Toronto, Canada: Association for Computational Linguistics, Jul. 2023, pp. 47--54. Available: \url{https://aclanthology.org/2023.trustnlp-1.5/}
\BIBentrySTDinterwordspacing

\bibitem{chowdhery2023palm}
A.~Chowdhery \emph{et~al.}, ``Palm: Scaling language modeling with pathways,'' \emph{Journal of Machine Learning Research}, vol.~24, no. 240, pp. 1--113, 2023.

\bibitem{golchin2023time}
S.~Golchin and M.~Surdeanu, ``Time travel in llms: Tracing data contamination in large language models,'' \emph{arXiv preprint arXiv:2308.08493}, 2023.

\bibitem{zhu2023dyval}
K.~Zhu \emph{et~al.}, ``Dyval: Graph-informed dynamic evaluation of large language models,'' \emph{arXiv preprint arXiv:2309.17167}, 2023.

\bibitem{ranaldi2024investigating}
F.~Ranaldi \emph{et~al.}, ``Investigating the impact of data contamination of large language models in text-to-sql translation,'' \emph{arXiv preprint arXiv:2402.08100}, 2024.

\bibitem{dong2024generalization}
Y.~Dong \emph{et~al.}, ``Generalization or memorization: Data contamination and trustworthy evaluation for large language models,'' \emph{arXiv preprint arXiv:2402.15938}, 2024.

\bibitem{shi2023detecting}
W.~Shi \emph{et~al.}, ``Detecting pretraining data from large language models,'' \emph{arXiv preprint arXiv:2310.16789}, 2023.

\bibitem{li2023avoiding}
Y.~Li \emph{et~al.}, ``Avoiding data contamination in language model evaluation: Dynamic test construction with latest materials,'' \emph{arXiv preprint arXiv:2312.12343}, 2023.

\bibitem{chandran2024private}
N.~Chandran \emph{et~al.}, ``Private benchmarking to prevent contamination and improve comparative evaluation of llms,'' \emph{arXiv preprint arXiv:2403.00393}, 2024.

\bibitem{jain2024livecodebench}
N.~Jain \emph{et~al.}, ``Livecodebench: Holistic and contamination free evaluation of large language models for code,'' \emph{arXiv preprint arXiv:2403.07974}, 2024.

\bibitem{riddell2024quantifying}
M.~Riddell \emph{et~al.}, ``Quantifying contamination in evaluating code generation capabilities of language models,'' \emph{arXiv preprint arXiv:2403.04811}, 2024.

\bibitem{pei2024betterv}
Z.~Pei \emph{et~al.}, ``Betterv: Controlled verilog generation with discriminative guidance,'' \emph{arXiv preprint arXiv:2402.03375}, 2024.

\bibitem{zhao2024codev}
Y.~Zhao \emph{et~al.}, ``Codev: Empowering llms for verilog generation through multi-level summarization,'' \emph{arXiv preprint arXiv:2407.10424}, 2024.

\bibitem{mankali2024rtl}
L.~L. Mankali \emph{et~al.}, ``Rtl-breaker: Assessing the security of llms against backdoor attacks on hdl code generation,'' \emph{arXiv preprint arXiv:2411.17569}, 2024.

\bibitem{alpaca}
R.~Taori \emph{et~al.}, ``Stanford alpaca: An instruction-following llama model,'' \url{https://github.com/tatsu-lab/stanford_alpaca}, 2023.

\bibitem{deepseekr1}
\BIBentryALTinterwordspacing
DeepSeek-AI, ``Deepseek-r1: Incentivizing reasoning capability in llms via reinforcement learning,'' 2025. Available: \url{https://arxiv.org/abs/2501.12948}
\BIBentrySTDinterwordspacing

\bibitem{claude37}
\BIBentryALTinterwordspacing
{Anthropic}, ``{Claude 3.7 Sonnet and Claude Code},'' 2025, accessed: 2025-02-28. Available: \url{https://www.anthropic.com/news/claude-3-7-sonnet}
\BIBentrySTDinterwordspacing

\bibitem{openaio3mini}
\BIBentryALTinterwordspacing
{OpenAI}, ``{OpenAI O3 Mini},'' 2024, accessed: 2025-02-25. Available: \url{https://openai.com/index/openai-o3-mini/}
\BIBentrySTDinterwordspacing

\bibitem{nie2025large}
S.~Nie \emph{et~al.}, ``Large language diffusion models,'' \emph{arXiv preprint arXiv:2502.09992}, 2025.

\end{thebibliography}

\end{document}